\documentclass[useAMS,usenatbib]{mn2e}
\usepackage{times}
\usepackage[dvips]{graphicx}
\def\mnras{MNRAS}

\title[Exact Equilibria of a Stellar System in a Linearised Tidal
  Field]{Exact Equilibria of a Stellar System in a Linearised Tidal Field}
\author[D.G.M. Mitchell and D.C. Heggie]{David G.M. Mitchell$^{1}$\thanks{E-mail:
david$\_$mitchell@bluebottle.com (DGMM); d.c.heggie@ed.ac.uk (DCH)} and Douglas C.
Heggie$^{1}$\footnotemark[1]\thanks{For Ken Freeman on his
  65th birthday}\\ 
$^{1}$School of Mathematics and Maxwell Institute for Mathematical Sciences, University of Edinburgh, King's
Buildings, Edinburgh EH9 3JZ, UK}
\begin{document}

\date{Accepted$\ldots$. Received$\ldots$; in original form$\ldots$}

\pagerange{\pageref{firstpage}--\pageref{lastpage}} \pubyear{2005?}

\maketitle

\label{firstpage}

\begin{abstract}
We study the motion of stars in a star cluster which revolves in a
  circular orbit about its parent galaxy.  The star cluster is
  modelled as an ellipsoid of uniform spatial density.  We exhibit two
  2-parameter families of self-consistent equilibrium models in which
  the velocity at each point is confined to a line in velocity space.
  We exhibit the link between this problem and that of a uniform
  rotating ellipsoidal galaxy.  With minimal adaptation, Freeman's bar
  models yield a third family.
\end{abstract}


\section{Introduction}
In a star cluster moving in a circular orbit about the axis
of symmetry of a galaxy, a star is subject to the self-gravity of the
cluster and the tidal field of the galaxy.  If studied in a rotating,
accelerating frame following the orbital motion of the cluster, it is
also subject to centrifugal and Coriolis forces.  If the spatial
distribution of stars is uniform within an ellipsoid, and if the usual
linear approximation of the tidal field is adopted, the
accelerations inside the ellipsoid are all linear in the spatial
coordinates.  There are then two normal modes of oscillation in planes
orthogonal to the axis of rotation.  \citet{fh} 
studied this problem in the case when one of the normal frequencies is
imaginary.  Taking for convenience the case of an ellipsoid of
revolution, they showed that it was possible to choose the axial ratio
so that it was equal to that of the orbits in the remaining normal
mode.  By a very simple orbit superposition it was possible to construct a
self-consistent model.

This problem closely resembles that of constructing a model of a
uniform, rotating ellipsoidal bar (Freeman 1966a,b,c; Hunter 1974,
1975).  This goal has been achieved in three cases: (i) elliptical
cylinders, in which the semi-major axis parallel to the axis of
rotation is infinite; (ii) so-called ``balanced'' systems, in which
the centrifugal and gravitational forces are equal along the major
axis; and (iii) disks with the surface density that is obtained by
projecting a uniform ellipsoid onto one plane.

In this paper we set out the connection between these two problems,
and give a more systematic treatment of the models which
Fellhauer \& Heggie stumbled upon.  In principle there are three
dimensionless parameters: two axial ratios, and one parameter which
measures the density of the system.  The requirement of choosing one
axial ratio appropriately reduces the family to a two-parameter
family.  Fellhauer \& Heggie chose to work with axisymmetric models,
which simplifies consideration to a one-parameter family.  
In cases where both normal frequencies are real, however, it is clear
that there is a possibility of a second two-parameter family.
Finally, when this second eigenfrequency is zero, the problem is
equivalent to that of Freeman's ``balanced'' bar.

We begin in the next section with a summary of the equations of motion
and their solution.  Then section 3 surveys the models based on a
single normal mode, and those which exploit the idea behind Freeman's
bar model.  The paper concludes with some final discussion, and an
Appendix describes the models in terms of distribution functions.

\section{Orbital Theory}

We use the rotating, accelerating frame in which the centre of the
ellipsoidal star cluster is at the origin, the $x$-axis points to
the centre of the galaxy, and the $y$-axis points in the direction of
orbital motion of the system.  (Fellhauer \& Heggie took the $y$-axis
in the opposite direction, but we switch in order to increase the
resemblance with the theory as set out by Freeman.)  Then the
equations of motion are
\begin{eqnarray}
  \ddot x +2\Omega\dot y + (\kappa^2-4\Omega^2)x &=& - A^2x\nonumber\\
  \ddot y -2\Omega\dot x &=& - B^2y\\
  \ddot z  + \nu^2z &=& - C^2z\nonumber,
\end{eqnarray}
where $\Omega$ is the angular velocity about the galaxy, $\kappa$ is
the epicyclic frequency, $\nu$ is the frequency of small motions
orthogonal to the galactic orbit of the cluster (if the cluster itself
is neglected), and $A^2,B^2,C^2$ are
coefficients of $x^2/2, y^2/2$ and $z^2/2$ in the expression for the
gravitational potential inside a uniform ellipsoid with semi-axes
$a,b,c$.  (They are functions of the density ($\rho$) and the axial
ratios $b/a, c/a$.)  For $a>b>c$, these expressions are given in \citet{f2}, eqs.(4)-(12),
which we would refer to henceforth as FII.4-12,  with corresponding
abbreviations for equations in his two
other papers.  Equivalent formulae will be found in Table 2-1 of
\citet{bt}, along with the special forms corresponding to
axisymmetrical and spherical cases.
Thus 
\begin{eqnarray}
  \ddot x +2\Omega\dot y + A_1^2x &=& 0\nonumber\\
  \ddot y -2\Omega\dot x + B_1^2y &=& 0\\
  \ddot z  + C_1^2z &=& 0\nonumber,
\end{eqnarray}
where 
\begin{eqnarray}
A_1^2 &=& \kappa^2-4\Omega^2+A^2,\label{eq:balance}   \\
B_1^2 &=& B^2, \mbox{~and}\\
C_1^2&=&\nu^2 + C^2
\end{eqnarray}
 (cf.  FI.8, FII.16).

For $\alpha\ne0$ the solutions of these equations are
\begin{eqnarray}
\label{eq:orbitx}
  x &=& A_\alpha\sin(\alpha t+\epsilon_\alpha) + A_\beta\sin(\beta
  t+\epsilon_\beta) \nonumber\\
\label{eq:orbity}  y &=& k_\alpha A_\alpha\cos(\alpha t+\epsilon_\alpha) - k_\beta
  A_\beta\cos(\beta t+\epsilon_\beta) \\
\label{eq:orbitz}z &=& A_\gamma\sin(\gamma t+\epsilon_\gamma)\nonumber
\end{eqnarray}
(FI.9, FII.17c), where $A_\alpha, A_\beta, A_\gamma$ are arbitrary
constants, which we can take positive, $\epsilon_\alpha, \epsilon_\beta, \epsilon_\gamma$
are arbitrary constants, $\gamma=C_1$, $\alpha, \beta$ are the two roots for $\xi$
of the quartic 
\begin{eqnarray}
\label{eq:frequencies}\xi^4 - (A_1^2+B_1^2+4\Omega^2)\xi^2+A_1^2B_1^2 &=& 0,  \\
k_\alpha &=& \displaystyle{\frac{A_1^2-\alpha^2}{2\alpha\Omega}} \\
\mbox{and\hskip 1.5truein}\label{eq:kbeta}k_\beta &=& \displaystyle{\frac{\beta^2 - A_1^2}{2\beta\Omega}}
\end{eqnarray}
(FI.10-11).  We adopt Freeman's convention that $\alpha<\beta$ when
both frequencies are real; when one frequency is imaginary (a case not
considered by Freeman) we denote the single real frequency by
$\beta$.  Then both $k_\alpha$ and $k_\beta$ are positive.  Note also
that, for a given galactic potential, they are functions of $\rho,
b/a$ and $c/a$.

\section{Three Families of Models}

\subsection{$\beta$ models}

We first summarise the family of self-consistent models which
generalises those found by \citet{fh}.  We take $A_\alpha=0$ in
eqs.(\ref{eq:orbitx}).  The models are then built
from the normal mode with frequency $\beta$, and so we refer to them
as $\beta$-models.   We can construct
self-consistent models provided that $k_\beta = b/a$.  When the
frequencies $\beta$ and $\gamma$ are incommensurable, each orbit fills the curved surface of an elliptic
cylinder.  The upper end of this cylinder is the ellipse
$x^2+y^2/k_\beta^2 = A_\beta^2, z = A_\gamma$, which lies on the
surface of the ellipsoid $x^2/a^2+y^2/b^2+z^2/c^2 = 1$ if
$A_\beta^2/a^2 + A_\gamma^2/c^2 = 1$.  Thus in general 
\begin{equation}\label{eq:agammamax}
 A_\gamma^2 \le A_{\gamma,max}^2 =
c^2(1 - A_\beta^2/a^2),   
\end{equation}
provided that $A_\beta^2\le a^2$.

For given $A_\beta, A_\gamma$ the distribution of $z$ is $f(z\vert
A_\gamma) = \displaystyle{\frac{1}{\pi\sqrt{A_\gamma^2-z^2}}}, \vert
z\vert\le A_\gamma$.  Given $A_\beta$, we choose the distribution of
$A_\gamma$ according to 
\begin{equation}\label{eq:fagamma}
f(A_\gamma) = \displaystyle{
\frac{A_\gamma}{A_{\gamma,max}\sqrt{A_{\gamma,max}^2 - A_\gamma^2}}} (0<A_\gamma<A_{\gamma,max})
\end{equation}
(For the case $a=c$, considered by Fellhauer \& Heggie, this is
equivalent to their eq.(19).)  With this choice we see easily that $z$
is uniformly distributed in $\vert z\vert\le A_{\gamma,max}$.  Finally, in order to generate a spatially
uniform density, we choose the distribution of $A_\beta$ so that
$f(A_\beta)dA_\beta$ is proportional to the mass inside the cluster in the cylindrical volume
between $x^2 + y^2/k_\beta^2 = A_\beta^2, (A_\beta + dA_\beta)^2$,
i.e.  
\begin{equation}\label{eq:fabeta}
f(A_\beta) =
\displaystyle{\frac{3}{a^2}A_\beta\sqrt{1-A_\beta^2/a^2}}, 
A_\beta < a
\end{equation}
 (cf. Fellhauer \& Heggie, eq.(18)). 
While these formulae are convenient for constructing a model, we also
(section \ref{sec:betadf}) provide a more conventional definition in terms
of a distribution function in phase space.

  Our purpose now is to
sketch the range of models which satisfy the essential requirement
$k_\beta = b/a$.  Since $k_\beta$ is a function of $\rho, b/a$ and
$c/a$, this equation may be solved for $\rho$ as a function of $b/a$ and $c/a$.  This can be exhibited by  plotting contours of $\rho$ in the plane
$b/a,c/a$.   Results of a numerical survey are shown in Fig.1,
where for definiteness we consider the case of an isothermal galaxy
potential, in which $\kappa = \sqrt{2}\Omega$.

\begin{figure}
\includegraphics[width=5.7cm,angle=-90]{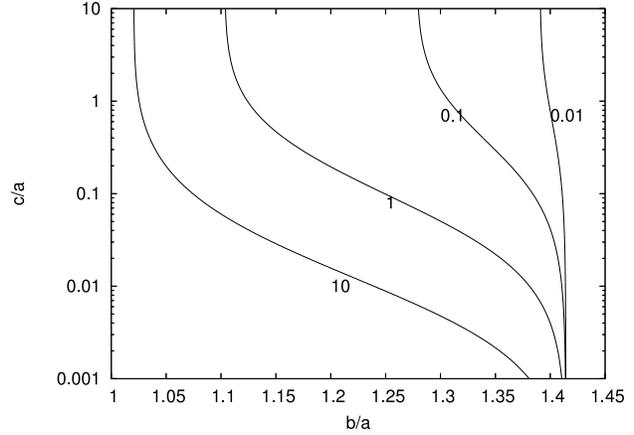}
  \caption{Contours of equal density for the $\beta$-models.  The value against each contour line is $G\rho/\Omega^2$ }
\end{figure}

Two  limiting cases can be discerned:
\begin{enumerate}
  \item {\sl Low density:}  If $\rho=0$ then $\beta = \kappa$ and
  $k_\beta = \sqrt{2}$, by eqs.(\ref{eq:frequencies}),
  (\ref{eq:kbeta}).  This is the limit at the extreme right of Fig.1.
\item {\sl Axisymmetric models:} If $\rho$ is very large and
  $a=b$, we have $A=B, \beta\simeq A+\Omega$ and $k_\beta\simeq1$, by
  the same pair of equations.
  This limit is reached at the extreme left of Fig.1.
\end{enumerate}

The models discussed by \citet{fh} lie on
the line $c/a=1$ in Fig.1.  (This subfamily is also depicted in Fig.4.)   We now see that they are part of a
two-parameter family.

\subsection{$\alpha$ models}

  For sufficiently low densities, the value
of $\alpha^2$ is negative, and orbital motions are unstable.  For
higher densities, however,  $\alpha^2\ge0,$ and so we have another family of self-consistent
models where now $A_\beta = 0$ and $k_\alpha = b/a$.  These
we refer to as $\alpha$ models.  Their distribution in parameter space
is shown in Fig.2.  Again there are axisymmetric models of arbitrarily
high density.  The construction of these
models is exactly analogous to that of $\beta$-models.
Again the subfamily on which $c/a=1$ is depicted in Fig.4.

\begin{figure}
\includegraphics[width=5.7cm,angle=-90]{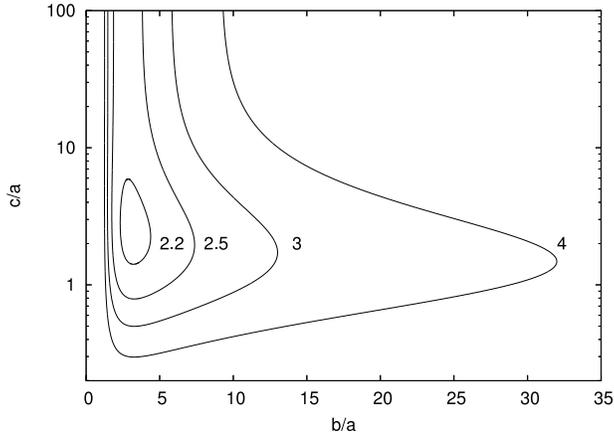}
  \caption{Contours of equal density for the $\alpha$-models, with labelling as in Fig.1.}
\end{figure}

\subsection{Freeman's Bars}

In F.II, Freeman described a series of three-dimensional ellipsoidal models of
uniform density for the case in which  $A_1 = 0$.  In the
galactic context this was referred to as the condition for a
``balanced'' ellipsoid.  For star clusters the corresponding
interpretation is that of a system which is marginally stable against
tidal disruption, as $\alpha = 0$.  They appear to exist only for
$G\rho/\Omega^2>0.3$ approximately, and their distribution is shown in Fig.3.

\begin{figure}
\includegraphics[width=5.7cm,angle=-90]{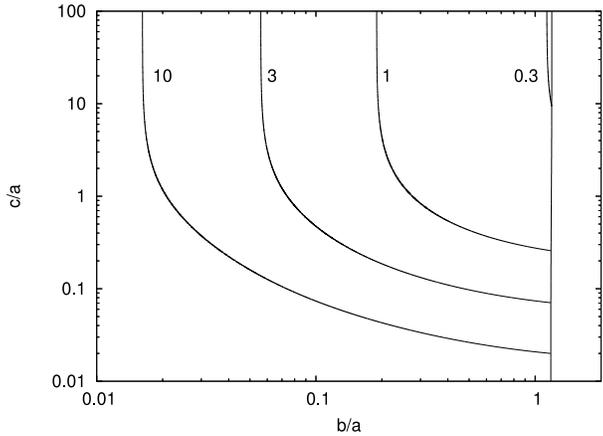}
  \caption{Contours of equal density for Freeman's Bars, labelled as in Fig.1.  The series ends at the near-vertical curve on the right, where  $k_\beta = b/a$.}\label{fig:freeman}
\end{figure}

\begin{figure}
\includegraphics[width=5.7cm,angle=-90]{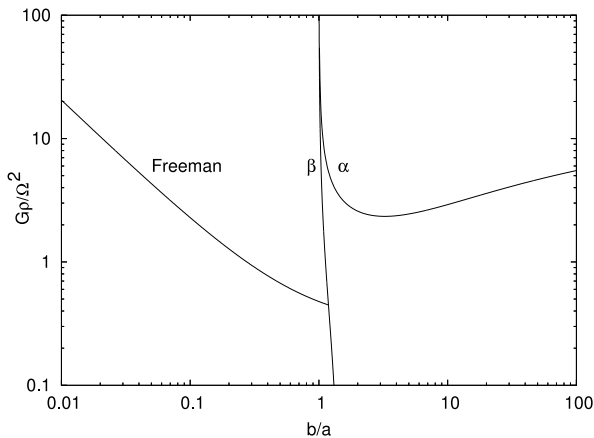}
  \caption{A cross section through all three families for $c/a=1$.   }\label{fig:all}
\end{figure}

For the case considered here, eqs.(\ref{eq:orbitx}a,b) are replaced by 
\begin{eqnarray}\label{eq:x}
  x &=& A_\alpha + A_\beta\cos(\beta
  t+\epsilon_\beta) \\
\label{eq:y}
  y &=& k_\beta
  A_\beta\sin(\beta t+\epsilon_\beta) 
\end{eqnarray}
(FII.17, with some changes of notation), where now
$\beta^2=4\Omega^2+B_1^2, k_\beta =\beta/(2\Omega)$, and the constant term $A_\alpha$ represents the zero-frequency
mode.  The effect of this term is that epicycles can be placed with guiding
centre at any point along the $x$-axis.  This freedom allows the
construction of models with an axial ratio different from that of the
epicycles.  A little thought, however, shows that the axial ratio of
the epicycles ($k_\beta$) must exceed that of the cluster ($b/a$) in
order that all parts of the cluster are accessible by epicycles lying
entirely within it.  This condition (which is best thought of as an
upper limit to $b/a$) gives the near-vertical curve at about $b = 1.2$
in Fig.\ref{fig:freeman}; realisable models must lie to the left of
this line.  This curve represents the intersection of the families of
Freeman and beta models, as can be seen in the cross section in
Fig.\ref{fig:all}.

\begin{figure}
\includegraphics[width=8cm]{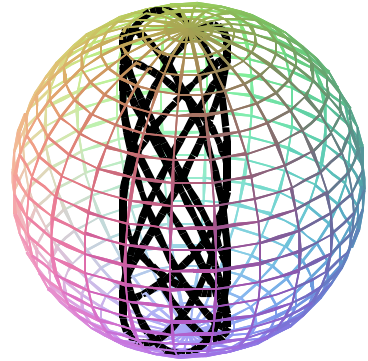}
  \caption{A typical orbit in the spherical Freeman model.  }\label{fig:orbit}
\end{figure}

An attractive special case is the spherical model $a=b=c$.  Then $A =
B = C = 4\pi G\rho/3$, and so eq.(\ref{eq:balance}), with $A_1=0$,  yields
$\displaystyle{\rho = \frac{3}{4\pi G}(4\Omega^2 - \kappa^2)}$.  A
typical orbit is shown in Fig.\ref{fig:orbit}.

For Freeman models it was found convenient to generate initial
conditions from the distribution function, which is described in FII.

\section{Conclusions and Final Remarks}

In this paper we have considered the problem of constructing an exact
self-consistent equilibrium of a star cluster in a tidal field.  The
usual linear approximation of the tidal field is adopted, and we
assume that the motion of the cluster in the galaxy is uniform and
circular.

We
have found three distinct families of models, in all of which the
cluster is an ellipsoid of uniform density.  Each is a two-parameter
family, the two parameters being the axial ratios of the ellipsoid.
Additional free parameters are the angular frequency of the motion
around the galaxy and the radius of the cluster.

All three models are built of epicyclic motions, of the kind which are
familiar in galactic dynamics, though the details of the orbits are
modified by the action of the cluster potential.  In two families of
models, one of which was partially explored by \citet{fh}, the space
density is built up by epicycles whose guiding centre lies at the
centre of the cluster and whose axial ratio (in the $x,y$ directions)
coincides with that of the ellipsoid.  We call these $\alpha$- and
$\beta$-models.  In a third family the guiding
centre of the epicycles may be displaced at various points along the
$x$-axis, which points to the galactic centre.  This family closely
resembles Freeman's family of ``balanced'' models for barred
galaxies \citep{f2}.  One of the families has models for all values of
the space density (for suitable choice of axial ratio), while the
other two exist only for densities above a certain minimum value.

We have given enough details of the distribution functions associated
with these models to allow explicit $N$-body realisations of them to
be constructed. 
Code is available from the authors for the
construction of the $\alpha$- and $\beta$-models, and for the
spherical Freeman model.

An interesting extension of the study of
these models would be consideration of their dynamical evolution,
under both collisionless and collisional processes.  For a
one-dimensional subfamily of the $\beta$-models this was considered by
\citet{fh}, with particular emphasis on low-density models.  They
found that the models are unstable, and that the time scale of the
instability increases as the density decreases.  Though they did not
explore high-density models in such detail, their results indicated
that the lifetime increases again at the highest end of the range they
studied.  

The question arises whether it is possible to construct more general
models along these lines.  Though no further analytic models have been
found yet, it has been pointed out to us by J.P. Ostriker that
this problem seems well suited to Schwarzschild's orbit-superposition
method \citep{sch}.

\subsection*{Acknowledgments}

It is a pleasure to thank R.H. Miller for drawing to our attention the
possibility that the $\beta$-models might be related to Freeman's bar
models.  DGMM acknowledges financial help from The University of
Edinburgh School of Mathematics through the award of a scholarship
from the William and Isabella Dick Bequest.  This paper was completed
during a visit to the Institute for Advanced Study under the
auspices of the Program in Interdisciplinary Studies, and DCH thanks
Piet Hut for his hospitality.  We thank the referee for helping us to
improve the presentation.

\appendix

\section[]{Distribution Functions}

\subsection{Integrals of the motion}

We invoke Jeans' Theorem and express the distribution function as a
function of the isolating integrals of motion.  As shown in FI and FII
these are
\begin{eqnarray}
\label{eq:ealpha} 
  E_\alpha &=& \frac{\alpha k_\alpha}{2\sigma}A_\alpha^2 =
  \frac{\sigma\alpha k_\alpha}{2}(\dot y - \beta k_\beta x)^2 +
  \frac{\sigma\alpha}{2k_\alpha}\left(\dot x + \frac{\beta y}{k_\beta}\right)^2\nonumber\\
\label{eq:ebeta}  E_\beta &=& \frac{\beta k_\beta}{2\sigma}A_\beta^2 =
  \frac{\sigma\beta k_\beta}{2}(\dot y + \alpha k_\alpha x)^2 +
  \frac{\sigma\beta}{2k_\beta}\left(\dot x - \frac{\alpha y}{k_\alpha}\right)^2\\
\label{eq:egamma}  E_\gamma &=& \frac{1}{2}\gamma^2A_\gamma^2 = \frac{1}{2}\dot z^2 +
  \frac{1}{2}\gamma^2 z^2,\nonumber
\end{eqnarray}
where $\displaystyle{\frac{1}{\sigma} = \alpha k_\alpha + \beta
  k_\beta}$.  These are in fact the energies of the three normal
  modes, and the total energy\footnote{in the rotating frame, so this is
  really the Jacobi integral} is $E = E_\alpha +E_\beta +E_\gamma$.

By combining results in FI.41 and FII.27, we see that the condition
that the envelope of this orbit just touches the surface of the model
is $J = 1$, where, in general,
\begin{equation}
\label{eq:j}
J = 2\sigma(k_\beta^2-k_\alpha^2)\left\{\frac{E_\alpha}{\alpha
  k_\alpha(a^2k_\beta^2 - b^2)} + \frac{E_\beta}{\beta
  k_\beta(b^2 - a^2k_\alpha^2)}\right\} + \frac{2E_\gamma}{\gamma^2c^2}.
\end{equation}
This expression is very useful for constructing distribution functions
vanishing outside the surface of the model.  When $\alpha$ is
imaginary only the $\beta$-models exist, and then $E_\alpha = 0$.
When $\alpha = 0$ we have the case of Freeman models, and then we find
somewhat different expressions for $J$ and the quantities it contains
(FII).  In general, FI.12--19 are a useful
resource of handy identities for checking some of the expressions in
the following subsections.

\subsection{$\alpha$- and $\beta$-models}\label{sec:betadf}

We consider here only the $\beta$-models, as the distribution
function for $\alpha$-models can then be constructed by obvious
substitutions.  When $A_\alpha = E_\alpha = 0$ and $b/a = k_\beta$,
eq.(\ref{eq:j}) becomes simply 
\begin{equation}
\label{eq:jbeta}
J = \frac{2\sigma E_\beta}{\beta a^2k_\beta} +
\frac{2E_\gamma}{\gamma^2c^2}.
\end{equation}

Our search for the distribution function can be guided by
eq.(\ref{eq:fabeta}).  Substituting the first part of
eq.(\ref{eq:ebeta}b) into eq.(\ref{eq:agammamax}), we see that
$A_{\gamma,max}^2 = \displaystyle{c^2\left(1-\frac{2\sigma}{a^2\beta
    k_\beta}E_\beta\right)}$.   From the first part of
eq.(\ref{eq:ebeta}c), and eqs.(\ref{eq:fagamma}) and (\ref{eq:jbeta}),
it follows that the probability density function of $E_\gamma$ is
$f(E_\gamma) = \displaystyle{\frac{1}{\gamma^2A_\gamma}f(A_\gamma)
  \propto \frac{1}{\sqrt{1-J}}}$.  
  In fact it can be seen that this expression, with
a $\delta$-function to set $E_\alpha=0$ and with suitable
normalisation, gives the full distribution
function:
\begin{equation}
  f(x,y,z,\dot x,\dot y,\dot z) = \frac{\rho\sigma\alpha}{2\pi^2\gamma
  c}\frac{\delta(E_\alpha)}{\sqrt{\displaystyle{1 - \frac{2\sigma E_\beta}{\beta a^2k_\beta} -
\frac{2E_\gamma}{\gamma^2c^2}}}}.
\end{equation}
It is easy to check that this yields the correct density, by
integrating over the velocity:  for the integration with respect to
$\dot x$ and $\dot y$ we convert to suitable polar coordinates, which
leads to an integral of the form
$\displaystyle{\int_0^\infty\delta(r^2)2rdr}$, and for this we adopt the
value 1.

\bsp

\label{lastpage}


\begin{thebibliography}{99}

\bibitem[Binney \& Tremaine(1987)]{bt} Binney, J., \& 
Tremaine, S.\ 1987, Galactic Dynamics. Princeton University Press, Princeton 

\bibitem[Fellhauer \& Heggie (2005)]{fh} Fellhauer, M., Heggie, D.C.\ 2005, 
A\&A, 435, 875 

\bibitem[Freeman (1966a)]{f1} Freeman, K.~C.\ 1966a, \mnras, 
133, 47 

\bibitem[\protect\citeauthoryear{Freeman}{1966b}]{f2} Freeman, K.~C.\ 1966b, \mnras, 
134, 1 

\bibitem[Freeman (1966c)]{f3} Freeman, K.~C.\ 1966c, \mnras, 
134, 15 

\bibitem[Hunter (1975)]{hunter75} Hunter, C.\ 1975, in Hayli A., ed,
  Proc. IAU Symp.~ 
69,  Dynamics of Stellar Systems.  D. Reidel, Dordrecht,  p.195 

\bibitem[Hunter (1974)]{hunter74} Hunter, C.\ 1974, \mnras, 166, 
633

\bibitem[Schwarzschild (1979)]{sch} Schwarzschild, M.\ 1979, 
ApJ, 232, 236 
\end{thebibliography}
\end{document}